\documentclass[12pt,manuscript]{aastex}
\shorttitle{HD 28867}
 
\begin{document}

\title{Deconstructing HD 28867}
\author{Frederick M.\ Walter}
\affil{Department of Physics and Astronomy, Stony Brook University,
Stony Brook NY 11794-3800\\fwalter@astro.sunysb.edu}
\author{Tracy L. Beck}
\affil{Gemini Observatory, 670 N. A'ohoku Pl., Hilo HI 96720\\
 tbeck@gemini.edu} 
\author{Jon A. Morse}
\affil{CASA, University of Colorado, Boulder CO 80309\\
       morsey@casa.colorado.edu}
\author{Scott J. Wolk}
\affil{Center for Astrophysics, 60 Garden St., Cambridge MA 02138\\
       swolk@cfa.harvard.edu}

\begin{abstract}
The 3\arcsec\ pair of B9 stars, HD~28867, is one of the brightest X-ray
sources in the Taurus-Auriga star forming region. In this multi-wavelength
study, we attempt to deduce the source of the X-ray emission.
We show that the East
component is the X-ray source. The East component has a near-IR excess
and displays narrow absorption lines in the optical, both of which are
consistent with a cool stellar companion. This companion is one
of the brightest low mass pre-main sequence stars in Tau-Aur; at
2~$\mu$m it and the B9 star are equally bright.
We see evidence for radial velocity variability in the cool component
of $>$34~km~s$^{-1}$.
It is not visible in K band speckle imaging,
which constrains the companion to lie within 14 AU of the B star.

 We also 
report on a possible fourth member of the group, an M1 star 18\arcsec\
south of HD~28867.
\end{abstract}
 
 
\section{Introduction}

The 3\arcsec\ visual pair HD~28867 (HR~1442; SAO 94002)
is one of the brightest
stellar X-ray sources in the Taurus-Auriga star formation region. 
Hoffleit (1982) 
lists the spectral type as B9~IVn.
Late~B stars are not usually bright X-ray sources. That this system is a
member of an active star forming region suggests that either young B stars
may be bright X-ray sources in some cases, or that the system may hide a
less massive but very active pre-main sequence star.

The Hipparcos catalog (ESA 1997) 
quotes $V=6.26$ and $\Delta V=0.05$~mag.
The individual $V$ magnitudes are 6.99 and 7.04.
The position angle of the pair is 277.1$^\circ$, with a separation of 
3.078\arcsec. 
The two stars, which we refer to as the East and West components,
have a common proper motion, with no significant change in
position angle or separation seen in 113 years (Hoffleit 1982).
Worley \& Douglass (1997) make no mention of a change in position angle or
separation in 159 years of observations (1830-1989).
The East component is the primary.

HD~28867 is in the vicinity of the L1551 dark cloud,
an active part of the Taurus-Auriga star formation complex.
In a spectroscopic search for previously unknown weak emission line (wTTS)
or Herbig Ae stars, Feigelson \& Kriss (1983) failed to note any H$\alpha$
emission associated with this system.
The star was, however, detected as a bright X-ray source with the
Einstein Observatory Imaging Proportional Counter (IPC)
in observation sequences 867 and 10538.
Walter et al.\/ (1988) quoted a count rate of 0.12~c~s$^{-1}$.
The system was catalogued as
MS0430.6+1754 by Gioia et al.\/ (1990), 
with a count rate of 0.10~c~s$^{-1}$
and log($f_x/f_v$)=$-$3.62 (Stocke et al.\/ 1991).  
It was further detected in the Einstein slew survey 
(Elvis et al.\/ 1992)  
as 1ES0430+179 with an IPC count rate of 0.11$\pm$0.03~c~s$^{-1}$.
Schmitt et al.\/ (1990) 
found a coronal temperature of 
log(T)=7.07$\pm$0.08, with an absorption column log(N$_H$)=20.0. The
0.2-4.0~keV flux was 2.7$\times$10$^{-12}$~erg~cm$^{-2}$~s$^{-1}$.

Carkner et al.\/ (1996)    
reported detections of HD~28867 in 
ROSAT and ASCA observations of the L1551 region.
Their spectral fit of the ROSAT PSPC data finds
kT=1.1~keV, log(n$_H$)=20.6, and
$f_x$= 4.1$\times$10$^{-12}$~erg~cm$^{-2}$~s$^{-1}$ (0.3-2.0~keV), with
comparable results (kT=1.2~keV, log(n$_H$)=20.6, and
$f_x$= 2.2$\times$10$^{-12}$~erg~cm$^{-2}$~s$^{-1}$ [0.8-5.0~keV])
from the ASCA
data. They show that the X-ray source is variable at $>$95\% confidence.
Bergh\"ofer, Schmitt, \& Cassinelli (1996) 
show that HD~28867 is one of the most X-ray luminous late-B
stars in the Yale Bright Star Catalog (Hoffleit 1982).   

White, Pallavicini, \& Kundu (1992) 
detected HD~28867 with a 6~cm flux of 0.36~mJy in a radio
survey of selected PMS stars. They were unable to establish which star in the
binary system was coincident with the radio source.
The radio flux is
comparable to that of radio-bright low mass PMS stars in Tau-Aur, and they
speculated that the source might be an unseen companion.

This system is IRAS point source 04306+1754. It was detected at 12$\mu$m with
a flux of 0.48~Jy; upper limits on the flux at 25 and 60$\mu$m are
0.58 and 0.40~Jy, respectively. The system was not resolved by IRAS.

Walter \& Boyd (1991) 
showed that HD~28867 shares the space
motion and parallax of the Tau-Aur star forming region.
The Hipparcos catalog (ESA 1997) 
provides a parallax of 7.71~mas (milli-arcseconds),
for a distance of 130$\pm$23~pc. HD~28867
appears to be
an intermediate mass member of the Tau-Aur association, on or near
the zero-age main sequence with an age of no more than a few million years.
At this distance the X-ray luminosity is about 10$^{31}$~erg~s$^{-1}$, which
makes it one of the most X-ray
luminous stars associated with the Taurus-Auriga complex.
Only HD~283572 (Walter et al.\/ 1987) 
is more luminous in the Damiani et al.\/ (1995) 
compilation
of Einstein IPC observations of Tau-Aur; none of the T~Tauri stars in the
ROSAT survey (Neuh\"auser et al.\/ 1995) 
exceed this luminosity.
That it is a bright and variable X-ray source, a radio source, and
is detected by IRAS at 12$\mu$m, strongly suggests that HD~28867 is
more interesting than the typical B9 star. The glare of the bright
B stars may hide a lower mass companion, or the young B9 stars themselves
may be magnetically active as a consequence of extreme youth. Here we present
multiwavelength photometric and spectroscopic observations which clarify the
nature of HD~28867. 

\section{Late B Stars as X-ray Sources}

Along the main sequence, the mean $f_X/f_{bol}$ ratio reaches a minimum
among the late-B/early-A stars (e.g., Pallavicini et al.\/ 1981). This
is consistent with expectations, as these stars lack the strong winds
(and resulting shocks) of the most massive stars (but see Feigelson et al.\/
2002), and lack the deep
convection zones which generate and amplify the magnetic fields of the
cool stars (Daniel, Linsky \& Gagn\'e 2002 provide a more comprehensive
review of these issues).
While O stars follow the trend $f_X/f_{bol}\approx$10$^{-7}$, Cassinelli
et al.\/ (1994) show that $f_X/f_{bol}$ decreases monotonically from about
10$^{-7}$ at spectral types B0-B1 to a few$\times$10$^{-9}$ at spectral type
B3, among stars not known to be multiple.
Schmitt et al.\/ (1993) reported detections of three of six late~B (B7-B9)
primaries in wide visual pairs (resolved in the ROSAT HRI),
with log($f_X/f_{bol}$) between -6.1 and -4.9, but
Schmitt (1997) failed to detect
X-ray emission from any of the three A stars earlier than spectral type A7 
in a complete sample of stars within 13~pc 
(there are no B stars in this sample). So while a few late~B/early~A stars
not known to be binaries may be bright X-ray sources, most are not.

A simple way to add excess X-ray flux to late~B/early~A stars is to
provide a lower mass binary companion. Given the main sequence lifetimes of
these stars, the cool star is likely to be young and active.
Golub et al.\/ (1983) and Caillault \& Helfand (1985) were the first to
suggest that the
X-ray bright late~B/early~A stars are close binaries with an active cool star.
White et al.\/ (1992) suggested this very explanation for
the X-ray and radio activity in HD~28867.
This would not be the first example of a late~B/cool star binary in a star
forming region. Casey et al.\/ (1995) showed that the
X-ray emitting late~B~star TY~Cra in
the CrA star formation region is a triple system,
consisting of two lower mass convective stars in
addition to the B8-9 primary.
The X-ray flux is consistent with that expected from the brightest
low mass PMS stars, such as HD~283572 (Walter et al.\/ 1987),
which is approximately the same 
mass and spectral type as the TY~CrA secondary.

Close binary systems are common, and
on a statistical basis there is no evidence to support the hypothesis that
late~B~stars in star forming regions are intrinsic X-ray sources:
Caillault, Gagn\'e, \& Stauffer (1994) 
showed that the fraction of B~star X-ray sources in the Orion Nebula
region was just that expected to be close binaries with late-type
companions. Feigelson et al.\/ (2002) concur, and show that the X-ray
luminosities require F/G type companions, rather than K/M companions.
Jefferies, Thurston, \& Pye (1997) reach the same conclusion
in the NGC~2516 cluster (they also suggest that magnetic chemically peculiar
B and A stars are more likely to be detected as X-ray sources than are normal 
late-B and A stars, either because they are intrinsically brighter or because
they are more likely to have a binary companion).
Daniel et al.\/ (2002) report that two mid-B and A Pleiads with F-G companions
are detected in a CHANDRA  observations, while two A stars not known to
have cooler companions are not detected.

Based on luminosity functions and hardness ratios, Hu\'elamo et al.\/ (2000)
also argue that the bright, hard X-ray sources in unresolved Lindroos (1986)
pairs must be late-type companions.
Indeed, using diffraction-limited near-IR images, Hu\'elamo et al.\/ (2001)
resolved one of three late-B Lindroos systems into the B star and a
pre-main sequence K star.

The alternative hypothesis is that HD~28867, a young intermediate mass star,
may retain some vestige of primordial magnetic activity, a ``naked'' Herbig~Ae
star, so to speak. Zinnecker \& Preibisch (1994) 
show that X-ray luminosities of some of the
more extreme Herbig Ae stars, including HR~6000, MCW~10980, and Z~CMa, 
are comparable to that of HD~28867, but HD~28867 is not known to have any of
the other spectroscopic characteristics of the Herbig~Ae stars. Giampapa,
Prosser, \& Fleming (1998) argued that two mid-B stars
in the Pleiades-age cluster IC~4665 are intrinsic X-ray sources. These stars,
with spectral types B5IV and B6V, are hotter and more masive than the stars
in HD~28867.

HD~28867 exhibits a flux ratio $f_X/f_{bol}$~=~10$^{-4.9}$
(Bergh\"ofer et al.\/ 1996) 
when referenced to the combined light of both stars. This is clearly much
larger than expected for a single, non-chemically peculiar B9 star.

\section{New Data}

In Table~\ref{tbl-optpos} we present the positions of the visible stars, taken
from the ``Double and Multiples: Component Solutions'' section of the 
Hipparcos and Tycho catalogs. 
We determined absolute positions in the K-band image (Figure~\ref{irimage})
by using a
reference grid of three stars from the USNO A2.0 catalog (Monet et al.\/ 1998)
which were also detected in this image. The HD~28867 pair is unresolved  
in the USNO A2.0 catalog.
We measured the positions of the three components on the K~band
image, and corrected the coordinates of the South component
to the Hipparcos reference frame.
The South component is not included in
either the Hipparcos or USNO A2.0 catalogs.
We also show the position of
the radio source from White et al.\/ (1992). This position is epoch 1990.1.
We have applied 10 years of proper motion (from the Hiparcos catalog) and
precessed the coordicates to equinox J2000.
The radio source lies within 0.3\arcsec\ of HD~28867E.

\subsection{Near-IR Photometry}

We obtained  JHK images of the system using the CIRIM infrared
imager (Elston 1999)  
on the CTIO 1.5m reflector on 19 February 1997 (Figure~\ref{irimage}).
We observed using
5 raster positions, one at the center and four each offset by 15\arcsec.
The integration time was 0.4~seconds, with three coadds at each raster
position. The net exposure time is 6 seconds per filter. The data were
coadded using the DOCIRIM
software\footnote{http://www.astro.sunysb.edu/fwalter/CIRIM/cirim.html}.
The plate scale is 0.6\arcsec\ per pixel.

We observed Elias et al.\/ (1982) 
flux standards approximately hourly; the photometric
solution has an RMS scatter of less than 2\% at J and K, and 3.5\% at H.

We measured magnitudes using aperture photometry. 
The standards were extracted 
using a 20~pixel (12.5\arcsec) radius aperture. We generated aperture
correction tables using the standard stars. We extracted the the stellar fluxes
using 3~pixel (1.9\arcsec)
apertures centroided on the peak of the flux, and corrected for
the light outside the aperture. The background is extracted within an annulus
between 8 and 28 pixels. The stellar magnitudes are given in
Table~\ref{tbl-irphot}. We verified that this returned pretty good absolute
photometry by extracting the flux from the pair of B stars using a 20\arcsec\
aperture.

The X-ray column of 10$^{20.6}$~cm$^{-2}$ corresponds to A$_V$=0.21~mag;
the observed $V-K$=0.23~mag of the West component
requires A$_V$=0.25~mag if this B9 star has intrinsic colors of 0.
These values are consistent within the uncertainties;
we adopt A$_V$=0.25~mag, which corresponds to A$_K$=0.03~mag for R=3.1.
We assume the reddening is interstellar and that both components are equally
affected.
After correcting for this reddening,
HD~28867E shows a significant near-IR excess of 0.18~mag in $J-K$, and
a $V-K$ color of 0.6~mag.

In addition to the pair of B stars, four other objects are visible
in the near-IR images. The brightest of these
is 18\arcsec\ south of the B stars. The position is coincident
with that of a Chandra X-ray source (\S\ref{cxrs}).
The near-IR colors are consistent with an early
M spectral type; it could be a either T~Tauri star at a projected distance of
2300~AU from the B~stars, or a foreground active M dwarf at about 80~pc.
We call this HD~28867S.
The other three near-IR sources are relatively faint objects
in the USNO A2.0 catalog (Monet et al.\/ 1998).

\subsection{X-rays}
\subsubsection{ROSAT HRI}
  We obtained ROSAT HRI image RH201046 in an attempt to resolve the
system. This 2.7~ksec observation (not shown) reveals a single bright source
spatially coincident with the binary.
There are 491 counts within 10\arcsec\ of the centroid, at a
nominal position of 4$^h$33$^m$32.954$^s$ +18$^\circ$1\arcmin 3.42\arcsec\
(J2000). 
We rebinned the image to 0.5\arcsec\ pixels, and
fit the profiles with Gaussians in the X and Y directions.
Within the uncertainties
the profiles are identical, with FWHMs, respectively, of
3.6 and 3.4\arcsec. The image is noticeably skewed, however.
A two-dimensional Gaussian fit gives FWHMs of 3.0 and 3.9\arcsec, respectively,
along the minor and major axes. The major axis has a position angle of
49$^\circ$.
The FWHM of the HRI point spread function (David et al.\/ 1997), 
fit in a similar manner, is 2.9\arcsec. We
conclude that there is no evidence that the source is resolved. Since the
greatest elongation occurs along an unphysical position angle,
we conclude it is likely to be instrumental.

The image also reveals a weak X-ray source (about 20 counts)
about 15\arcsec\ south of the B stars. This is spatially coincident with
the near-IR source HD~28867S.

\subsubsection{Chandra HRC-I}\label{cxrs}

As a followup to the ROSAT HRI image, we obtained 
a 4680~sec Chandra HRC-I image (observation 200037). With its $\sim$0.5\arcsec\
resolution, this image (Fig.~\ref{hrci}) clearly shows
a single bright X-ray source.
The count rate is 0.33~c~s$^{-1}$. Adopting the Carkner et al.\/ (1996)
spectral model, this count rate corresponds to a flux of 
3.1$\times$10$^{-12}$~erg~cm$^{-2}$~s$^{-1}$ in the 0.3-10~keV band. This
flux lies within the range of previously observed fluxes.
There is no significant variability during this short observation.
The source position is coincident to within 0.22\arcsec\ with that of
HD~28867E (Table~\ref{tbl-optpos}).

There is no emission evident from the vicinity of West B9 component. Within a 
1.32\arcsec\ extraction circle, we detect 3.9$\pm$2.5 counts, for a 
count rate of $<$1.5$\times$10$^{-3}$ (3$\sigma$) . The limiting 
log($L_X/L_{bol}$) of $-$7.2 is consistent with normal stars of this 
spectral type.

A weak source was detected within 0.15\arcsec\ of the near-IR position of the
M star HD~28867S (Table~\ref{tbl-optpos}).
We detected 17 photons from this source, for a net count rate of
3.6$\pm$0.9$\times$10$^{-3}$~s$^{-1}$.
Assuming a 0.8~keV thermal plasma spectrum, and
correcting for A$_V$=0.25 mag of extinction,
the luminosity is about
1.5$\times$10$^{29}(D/130~{\rm pc})$~erg~s$^{-1}$,
and the $f_X/f_{bol}$ ratio is about -2.6. These are consistent with
expectations for an active M star.

\subsection{Ultraviolet Spectroscopy}

We obtained IUE (Boggess et al.\/ 1978) spectra of HD 28867 on 1987 March 10
as part of program TTJFW (Table~\ref{tbl-iue}).
Three observations, a low dispersion SWP spectrum,
and high dispersion SWP and LWP spectra, were taken with the target in the
large aperture. The pair is unresolved in the large aperture; the
spectra look like those of late-B stars.

We obtained followup SWP-HI
spectra in 1994, using the IUE small aperture to separate the
two components, as part of program BXPFW.
We first centered the star in the
FES, slewed to place the light centroid in the small aperture,
and then performed blind offsets of 
the telescope by +1.5\arcsec,-0.25\arcsec\
in RA, DEC for the East component, and by -1.65\arcsec,+0.25\arcsec\ for the
West component. 
The archival SWP~49847 spectrum is mis-calibrated as a large
aperture observation (the script correct labels it as a small aperture
observation). We compensated for this by multiplying the calibrated flux
by the small-to-large aperture ratio in the file {\it swpabscal.fit}.

The IUE small aperture has a 3.0 arcsec diameter, so there
may be some contamination from the other star, especially if there was some
small error in the centroiding or the offset slews. Based on the FES
offsets, the final separation between the two apertures on the sky is
2.7\arcsec,
which is less than the true binary separation.
The East component appears about twice as bright as the West component in the
extracted spectrum, which suggests that either the West component was not well
centered in the aperture, or that the East spectrum includes some light from
the West component. The summed flux is about half the flux in the large
aperture observation, SWP~31968.

The spectra, though noisy, show no gross abnormalities.
There are no emission lines, and the absorption lines are consistent with the
spectral types. The spectral shapes and line depths of the two stars are
very similar. We compared
Kurucz (1979) model atmospheres to the spectra in the
1200-1600\AA\ region in an attempt to pin down temperatures. For an assumed
log~g=4 and A$_V$=0.25, both stars are fairly well described by 10,500K
model stars. More detailed analysis is likely to yield unreliable results
due to the contamination of the spectra.

\subsection{Optical Spectroscopy}

We obtained high dispersion optical spectra with the echelle spectrograph
on the KPNO 4m Mayall telescope on two occasions. 
We obtained two spectra on 27 October 1988,
with about 1.2~arcsec seeing, some light cirrus at the start of the night,
and the moon 4 days past full in Auriga. We used a 1\arcsec\ slit oriented
E-W with a 4\arcsec\ decker length. With one star centered in the slit,
the other star was visible at the end of the decker. We extracted the spectra
by fitting two Gaussians at each wavelength coordinate, and taking the
integral of the Gaussian as the number of counts at that wavelength.
The TI2 CCD gave a wavelength coverage of about 5450 to 7200\AA.
Integration times were 50 and 60 seconds, respectively, for the E and W
components.

On 24 January 1997, under excellent seeing conditions, we reobserved the pair.
The moon was about 2 days past full, but was below the horizon at the time
of the observations.
We used the same slit and decker, but off-centered the target near the end
of the decker to minimize contamination from the companion. This simplified the
reductions. The TK2B CCD provided wavelength coverage from about 4300 to
7400\AA. Each star was observed for 5 minutes. Because we used a simple
boxcar extraction rather than an optimal extraction, the S/N is limited
by pixellation in the image. 

We determined the absolute observed wavelength scale using the telluric water 
vapor lines near H$\alpha$. We then applied the heliocentric correction
to establish the wavelength scale.

Segments of the spectra are shown in Figure~\ref{fig-optsp}. In each panel
the upper trace is the East component while the lower trace is the West
component (The central trace is discussed in \S\ref{sec-ss}).
The West component shows a rotationally-broadened B9 spectrum,
with superposed narrow telluric features.
The spectra of the East component show prominent broad absorption
lines unexpected in a late~B star, including Li~I~$\lambda$6707\AA, with an
equivalent width of 70$\pm5$~mA. The strongest of lines (in general, those
with W$_\lambda>$30~m\AA, but excluding broad blends) are tabulated in
Table~\ref{tbl-optabs}. Line identifications are made using the 
Wallace \& Hinkle (1998) Solar atlas, with wavelengths obtained from the
Kurucz \& Bell (1995) line list. 
The wavelengths presented in
Table~\ref{tbl-optabs} are the centroids of the broad and often blended
lines measured in the 1997 January spectrum,
corrected to heliocentric wavelengths.

Most of the lines are identified with \ion{Fe}{1} and
\ion{Ca}{1} lines prominent in solar-like stars.
The lines are broad. A fit to the Li~I line gives an equivalent 
V~$sin i$ of 65~km~s$^{-1}$. Broadening of the other lines is consistent with
this.

We measure the radial velocity of the cool star by cross-correlating the
spectum against a template made of the strong lines identified in the
spectrum. The radial velocity of the cool star is -5~km~s$^{-1}$ in the 1988
observation, and +29~km~s$^{-1}$ in the 1997 observation, with uncertainties 
estimated to be about 2~km~s$^{-1}$. Assuming a 
$\gamma$~velocity of 17~km~s$^{-1}$ appropriate for the Tau-Aur association,
the projected orbital velocity of the cool star must exceed about
22~km~s$^{-1}$.

\subsection{Radial Velocity Measurements}\label{secrv}

Both components of the HD~28867 system were observed with the Coud\'e feed
at Kitt Peak National Observatory on the nights of 9-10 December 1990
and 20 October 1991 as part of an extensive OB star radial-velocity
monitoring program. The observing, data reduction,
and radial-velocity measurement techniques are described in Morse, Mathieu, \&
Levine (1991). The configuration for the first run employed the RCA 512x512 CCD
detector with the Coude Spectrograph Grating B to provide a dispersion of
0.261 \AA\ pix$^{-1}$ over the wavelength range 3715\AA\ -- 3848\AA.
For the second run the TEK2k CCD was used with Grating B to provide a
dispersion of 0.112\AA\ pix$^{-1}$ over 3690\AA\ -- 3860\AA.
Each observation achieved S/N $\sim$ 40 per pixel. Both late-type and
early-type radial-velocity standard stars (e.g., Fekel 1985) were observed
in close temporal proximity to the HD~28867 pair observations.
 
The heliocentric radial-velocity measurements are based on cross correlations
of observed spectra with a grid of Kurucz synthetic template spectra that span
a broad range of effective temperature and $v$ sin$i$. The typical
measurement uncertainties reported by Morse et al.\ (1991) are $\sim$2 
km~s$^{-1}$,
though 1~km~s$^{-1}$ precisions are attained for sharp-lined (low-$v$ sin$i$)
standard stars, such as HR~1149 (B8~III), HR~2010 (B9~IV), and
HR~1389 (A2~IV). Their study found that using the Balmer series of lines
near 3700\AA\ provided robust results over the spectral type range late-O
to early-A for all $v$~sin$i$'s.
 
The cross correlation procedure ``selects" an appropriate template spectrum
by measuring the quality of the template match using the so-called R value
(see Latham 1985). The templates used for the HD~28867 pair measurements were
T$_{\rm eff}$ = 11,500 K, log g = 4.5, $v$ sin$i$ = 300~km~s$^{-1}$
for the west
component, and T$_{\rm eff}$ = 10,500 K, log g = 4.5, $v$ sin$i$ =
200~km~s$^{-1}$
for the east component. These stellar parameters should be regarded as
first-order estimates, limited by the density of the synthetic template grid,
but indicate that the east component is slightly cooler and a somewhat slower
rotator than the west component.
 
The heliocentric radial velocities derived for the west component were +16.3
and +15.6 km~s$^{-1}$ for the December 1990 observations and
+26.9 km~s$^{-1}$ for the October 1991 observation.
For the east component, the velocities were +10.3 and +11.9 km~s$^{-1}$
for December 1990 and +21.1 km~s$^{-1}$ for October 1991.
At first sight, one might conclude that both stars are radial-velocity
variables, each showing offsets from one observing run to the next that are
significantly larger than the nominal measurement precision. The three
standard stars mentioned above were observed close in time to the HD~28867
pair observations and show excellent ($\sim 1$ km~s$^{-1}$) repeatability.
However, during each run the two components are consistently
offset from each other by $\sim 5$ km~s$^{-1}$ and the run-to-run offsets for
each star are $\sim$ +10 km~s$^{-1}$.

This behavior makes us wary that there may be some unknown systematic problem
with the HD~28867 pair observations, despite the good behavior of the
standard stars.
We note that Abt \& Biggs (1972) 
and Hoffleit (1982) 
note that HR~1442 may be a radial velocity
variable. Two published values differ by 15~km~s$^{-1}$.
It is not clear whether these refer to the combined light of the pair, in
which case the interpretation is complicated.

\subsection{Speckle Imaging}

We obtained speckle images of HD~28867 on the evening of 2001 January 12 using
the facility IR camera ``NSFCam'' at the IRTF in non-photometric conditions.
We used the 0.055\arcsec~pix$^{-1}$ plate scale to obtain three hundred
0.078 second speckle exposures
in each of the J, H and K photometric bands.   We used a shift-and-add analysis
method to produce the final images.  Frames that had less than 30\% of the peak
flux in the best image were excluded as images that were taken in poor
instantaneous seeing conditions.  The J, H and K-band images were thus
constructed from centroiding and combining the best 169, 207 and 224 frames,
respectively.
 
The speckle image of HD~28867E in the K-band shows only a single source, with a
spatial profile indistinguishable from that of HD~28867W
(Fig.~\ref{fig-kspec}). To estimate limits at
which we can detect a companion to HD~28867E, we determined the sensitivity to
companions at a range of brightnesses and separations.  In
Figure~\ref{fig-ksens} we present
the sensitivity ($\Delta$K) versus angular separation as derived using
HD~28867W
as a point spread function calibrator. The sensitivity at the projected
separation from the central star is essentially the limiting magnitude at this
distance, ten times the standard deviation over the mean counts in each 2 pixel
wide annuli.  Based on the K-band speckle analysis, we can rule out with
confidence a late type companion with a K-band brightness within 1 magnitude of
that of the B star at a separation of more than 0.25\arcsec.
This corresponds to a projected separation of 32~AU.

A cool companion to HD 28867E will be brighter at the longer
wavelengths, so there could be a slight positional shift in the 
separation between the stars as a function of wavelength.
To test this, we measured the spatial separations between the two stellar
centroids. There is no significant difference between the separations
in the 3 bands, and no trends with wavelength.
The mean separation between the centroids in the J, H, and K
band images is 3.070$\pm$0.004\arcsec. 
The separation between the optical components from the Hipparcos measurements
is 3.078$\pm$0.004\arcsec\ (ESA 1997). 
There is no evidence for a significant shift
in the centroid position between the optical, where the light from a cool
component is insignificant, to the K band.

Formally, the optical and near-IR position angles are significantly
different, at a level of
1.00$\pm$0.14$^{\circ}$, corresponding to a northward shift of 0.98 pixels
in the East component. However, given
the accuracy to which the camera orientation is known,
this is probably not significant.

If we accept that the relative shift between the optical and near-IR
centroids is no
more than one 0.055\arcsec\ pixel, then we can put another 
limit on the maximum separation of the two stars. We show below
(\S\ref{sec-ss}) that any cool component is roughly as bright as the B star
in the K band. A shift in the centroid of $<$0.055\arcsec\ therefore
requires a 
projected separation $<$0.11\arcsec\ (14~AU). This is consistent with the
speckle sensitivity estimate shown in Figure~\ref{fig-ksens} for
$\Delta$K=0~mag.


\subsection{Near-IR Spectroscopy}
 
On 29 December 2001 we used the SPEX
(Rayner et al 1998; 2002) low resolution spectrograph in the short
wavelength, cross-dispersed (SXD) mode, to obtain a spectrum from 0.8 through
2.5~$\mu$m at a spectral resolution of about 1500. Conditions were
photometric.
We observed in AB mode,
with a 100\arcsec\ throw. We observed the brighter East component
for 8 minutes (30 second integrations, 2 coadds, 8 AB cycles), and the West
component for 12 minutes. We also observed the South component for 16
minutes. We extracted the spectra using the Spextool software version 2.1.
We flatten the spectra by assuming that the continuua of the
A0 stars follow F$_\lambda\propto\lambda^{-4}$ power laws.
We then merged the individual orders into a single spectrum.
We used the observed fluxes of the A0 stars, assuming $V-K$=0,
to correct to absolute fluxes. 

We derive a K magnitude of 6.79 for the West component from the SPEX
spectrophotometry. This is within 2\% of the photometric magnitude. The
derived K magnitude of 6.07 for the East component is 0.1~mag brighter
than the photometric magnitude (Table~\ref{tbl-irphot}). Since there
is no evidence for intrusion of light from the (fainter) West component
into the slit, the spectra were taken back-to-back, and the calibrations are
stable, this may be evidence of variability of the cooler
companion. Variability will be small in the optical, where the cool companion
contributes only a few percent of the flux.
 
Spectra of HD~28867E and W are shown in Figure~\ref{fig-sp}.
A spectrum of HD~28867S, scaled up by a factor of 10, is also shown in
Figure~\ref{fig-sp}. The West component is a good approximation to the
Rayleigh-Jeans tail of a blackbody, while the East component has excess
flux at long wavelengths. The South component is a good match to an M1 star.
This spectral type is consistent with the JHK colors.
 
The ratio of the flux in the East and West components is shown in the upper
trace in Figure~\ref{fig-sprat}. The emission in the H~I Paschen and Brackett
series is an artifact of the division process (the lines are proportionately
shallower in the East component, after dilution). The stars were observed
consecutively, at airmass less than 2, and most of the telluric absorption
features divide out. However, after division there are
apparent narrow absorption lines in the flux ratio. Inspection of the
spectra shows that they are in absorption in the East component. The
absorption features in the K~band are identified in
Figure~\ref{fig-expsprat}.
 
We identified the absorption features by comparison with a Solar IR
atlas (Livingston \& Wallace 1991).
The absorption lines are those of neutral species (\ion{Fe}{1},
\ion{Si}{1}, \ion{Mg}{1}, \ion{Na}{1}, \ion{Al}{1}),
as well as the CO $\Delta\nu$=2 bandheads. This
suggests a cool stellar component of spectral type G-K, 
in good agreement with the optical absorption line spectrum.
 
\section{Spectral Synthesis}\label{sec-ss}
 
On the assumption that HD~28867E is a composite of a late~B star
and a cool dwarf,
we used the near-IR spectrum and an atlas of near-IR spectral standards to
estimate the properties of the cool dwarf. We observed the near-IR spectral
standards (spectral types G1 through L2)
during the same observing run, and with the
same instrumental setup.
 
The radial velocity template and the IUE spectra suggest that HD~28867E is 
somewhat cooler than HD~28867W.
Therefore, we modelled the B star with a range of temperatures, 
from 1500K cooler than the West component to the same temperature.
We used the spectrum of the West component, scaled by the ratio of
blackbodies at the two temperatures, as the template for the B star component.
We then subtracted this B star model from the spectrum, 
and fit the difference spectrum,
F$_{\rm HD~28867E}-({\rm F}_{\rm HD~28867W} \times
(BB(T)/BB(11,500K))+{\rm F}_{\rm standard}$).
At each spectral type, we scale the flux of the standard to match the
excess flux, and then subtract the standard. Ideally, the residuals will be
uniformly zero for the correct choice of companion.

We determine a best fit companion spectral type from the colors of the
difference spectrum, by
a $\chi^2$ analysis of the residuals, and by fitting the slope
and curvature of the residuals. 
The best spectral type for a companion with dwarf colors ranges from 
early-G to mid-K,
depending on the assumed temperature of the B star;
more luminous companions will have earlier spectral types.
In Table~\ref{tbl-comp} we list the derived parameters of the companion
over the tested range of B star temperatures.
The spectral type is determined by comparison with the spectral energy
distributions of the IR standards. The radius is derived from the flux
normalization. 
The magnitudes in Table~\ref{tbl-comp} have been corrected for A$_V$=0.25~mag.
In addition, we present the ratios of the B star to companion flux in the
R$_C$ and K bands.  
The range of possible locations of the companion in the
H-R diagram is shown in Figure~\ref{fig-hrd}.

Note that this analysis considers only the broadband spectral energy
distribution, and not the information in the spectral lines.
We generated synthetic optical spectra by adding the spectrum of HD~28867W
(the B9 template) and G-K spectra diluted by the appropriate factor 
(see Table~\ref{tbl-comp}). Prior to adding the spectra, 
we added a Li~I line with an equivalent width corresponding to a cosmic 
abundance log~n(Li)=3.3 (Pavlenko \& Magazzu 1996), 
appropriate for a PMS star, and then
smoothed the spectrum to approximate a rotational broadening
$Vsin~i$=65~km~s$^{-1}$. The middle traces in Figure~\ref{fig-optsp} show
this synthetic spectrum for a G2 companion. 
It successfully reproduces the broad
spectral lines, including Li~I~$\lambda$6707\AA, in the spectrum of
HD~28867E. By visual inspection of these composite spectra, we were
able to exclude a companion as late in spectral type as K5, but we cannot
distinguish between a G2 and K0 companion. The increase in the spectral
dilution in the cooler stars almost exactly compensates for the deepening
lines, and the rapid rotation washes out all but the strongest features,
which are not strongly sensitive to temperature in this range

The lower trace in Figure~\ref{fig-sprat} shows
the ratio of the observed flux from HD~28867E to this synthetic spectrum
in the near-IR. The spectral energy distributions match, as do the 
lines in the K band. 
In particular, the $\Delta\nu=2$ CO bandheads and the
$\lambda$2.21$\mu$m Na~I doublet and $\lambda$2.26$\mu$m Ca~I triplet
go into emission (are undersubtracted) if we force the
companion spectral type later than about K0.

Greene \& Meyer (1995) and Hodapp and Deane (1993) have quantified
relations between various absorption line indices and T$_{eff}$ or
spectral type. Prior to measuring the lines, we accounted for telluric
absorption by dividing the spectrum by the normalized 
spectrum of an A0 star observed at the same air mass. 
The equivalent widths of the Na~I doublet and the Ca~I triplet
in the difference spectrum depend on the selected temperature differential.
Even after correcting for the telluric absorption, the
continuum is not well defined, and lines are blended, at this resolution.
The summed equivalent width (Greene and Meyer's atomic index) range from 
2.5\AA\ for a 1500K temperature differential to 3.0\AA\ for no
differential.
This corresponds to spectral type K0V to K2III, which is slightly cooler
than indicated by the broadband spectral energy distribution or the 
optical lines. The CO bandheads subtract well for late-G spectral types;
the gravity dependence of the CO bandheads permit stars as cool as
early-to-mid~K for luminosity class III
(see Figures~8 and 10 of Hodapp \& Deane 1993).

\section{Conclusions}

We conclude that this star does not support the premise that
that late-B stars can be intrinsically luminous X-ray sources.
HD 28867E is a composite of a late B star and a cooler PMS star.

This conclusion is based on optical and near-IR spectroscopy, as well as the
near-IR photometry. The spectroscopic data limit the spectral type of
the companion to between about G0 and K0, with the spread
due to the uncertainty of the temperature of the B star component. 
Uncertainties in the reddening cannot affect the shape of the near-IR
spectral energy distribution sufficiently to affect this conclusion.
One can refer to the location of the companion in the H-R diagram 
(Fig.~\ref{fig-hrd}) to draw further conclusions. There we show the
location of the star referenced to evolutionary tracks and isochrones
from Baraffe et al.\/ (2002) and D'Antona \& Mazzitelli (1994).
The Baraffe et al.\/ tracks use a mixing length equal to the pressure scale
height and Y=0.275; the D'Antona \& Mazzitelli tracks are those using
mixing length convection and Alexander opacities.
Note that the tracks and isochrones differ. 
The Baraffe et al.\/ tracks incorporate more recent input physics,
but are lacking above 1.4~M$_\odot$. From the Baraffe et al.\/ tracks,
we can conclude that the mass exceeds about 1.2~M$_\odot$, and that the
age, at least at the red end of the allowable range, is 1-3~Myr,
consistent with the age of the Tau-Aur T~association. The D'Antona
and Mazzitelli tracks permit a younger and less massive star.
For reference, the open circle shows the location of
HD 283572, one of the brightest low mass PMS stars in Taurus.

For a temperature difference of 1000K,
the spectrophotometric deconvolution leaves a hot component with
$V$=6.92 and $V-K$=0.03 (unreddened).
This star is 0.13~mag fainter than the West component.
Using the simple relations that the $V$~band flux will be proportional to
R$^2$T, that R$\sim$M, and that T$\sim$M$^3$, where R, T, and M are 
respectively the
stellar radius, mass, and temperature, we expect a star 1000K cooler than the
West component to be about 14\% fainter. This consistency adds confidence
to our spectrophotometric deconvolution.

The X-ray and radio emission is fully consistent with this interpretation
of the system as a B star and a cooler companion.
The IRAS 12$\mu$m flux is inconsistent with this, as it overestimates
the flux, extrapolated from K by $\lambda^{-4}$, by a factor of 4.
This could be attributable to a
circumstellar disk around one or both of the stars. The disk would need to 
have a substantial inner hole, as it does not appear to contribute
significantly in the K~band.

With the data in hand, we can start to constrain the orbit of the system
(assuming this is no chance projection).
The cool star is brighter than the B star in the K band so long as the
spectral type is earlier than K0 (Table~\ref{tbl-comp}). The 14~AU limit
from the K band imaging is a limit for the semi-major axis for circular
orbit. The system is unlikely to be oriented pole-on.
The $\sim$65~km~s$^{-1}$ line width 
constrains the rotation period to be less than 2.0~(R/2.6~R$_\odot$)~days.
This is consistent with
the rotation periods of the more massive of the low mass PMS stars, and
suggests that the inclination $i$ exceeds about 30$^\circ$ (corresponding to
a 1~day rotation period). If the rotation and orbital axes are coaligned, this
$i$ applies to the binary system as well.
No evidence has been reported for an optical
eclipse, so the inclination must be less than 90$^\circ$, but this is not a
strong constraint since the semi-major axis is likely to be large compared to
the stars.

A B9 star has a mass of about 3~M$_\odot$. It is unfortunate that the
B star radial velocity measurements are inconclusive, since for reasonable
secondary masses ($>$0.5~M$_\odot$)
the radial velocity of the B star should vary by 
$>$4~km~s$^{-1}$. For orbital inclinations $i>$30$^\circ$, and
0.5$<$M$_2$/M$_\odot<$2.0, the semi-major axis of the system lies between
about 0.8 and 4.7~AU (6 -- 36~mas),
and the orbital period lies between about 4 months and
5.5~years. All these values are overestimates, given that we only have
a lower limit on the orbital V~$sini$ of the secondary.

Further observations will throw more light on this system.
We have renewed our efforts to determine the radial velocities
of the B stars in the HD~28867 pair
using the echelle spectrograph at the ARC 3.5-m telescope.
High dispersion spectra in the red and near-IR can be used to determine the
orbit of the cool companion, and high
dispersion, high S/N spectra in the 
near-IR will yield better limits on the spectral type. 
Diffraction-limited K-band imaging on a larger telescope may be able to 
directly resolve the system. We have hopes that in the near future this
system will be one of the few astrometric and spectroscopic binaries among
the pre-main sequence stars, and hence an important calibrator for stellar
evolution models.

\acknowledgments
We acknowledge use of the SIMBAD databases, maintained by the Centre de
Donn\'ees Astronomiques de Strasbourg (CDS). We made use of the
catalogs maintained at the former Astronomical Data Center at NASA
Goddard Space Flight Center. We are grateful to the staffs of
the Kitt Peak National Observatory, the NASA Infrared Telescope Facility, and
the IUE, ROSAT and Chandra observatories for their assistance in obtaining
data.
This research was funded in part
by ROSAT grant NAG5-1594 and Chandra grant GO0-1077X
to Stony Brook University.

\clearpage

\figcaption{The inner 1\arcmin\ of the CIRIM K band image.
HD~28867 is the bright pair in the center.
HD~28867S, an early M star, is 15\arcsec\ to the south.
The scaling is linear.
\label{irimage}}  

\figcaption{The central part of the Chandra HRC-I image.
The data have been binned into 0.26\arcsec\ pixels.
The three circles mark the relative positions of the three stars.
The circles have radii of 1\arcsec.
The axes are labeled in arcsec from HD~28867E.
\label{hrci}}  

\figcaption{Segments of the optical spectra. In each panel, the upper
trace is the spectrum of HD~28867E, the lower trace is the spectrum of
HD~28867W, and the thick trace in the center is the composite spectrum
made by summing the spectrum of HD~28867W with an appropriately scaled
and broadened spectrum of the G2V star HD~211476. A
Li~I $\lambda$6707\AA\ line corresponding to log~n(Li)=3.3 has been
added. All spectra are normalized to their median; the upper and lower
spectra in each panel are offset by $\pm$0.1 units, respectively.
The scalloping visible in the HD~28867W spectrum, prominent in the
upper panel, is an artifact of the data extraction.
The narrow lines near $\lambda$6475 are telluric H$_2$O.
\label{fig-optsp}}  

\figcaption{The speckle image in the K band from the IRTF.
The scaling is logarithmic above a level of about 3~times the sky.
The stellar cores are sharp, and the image is
clearly diffraction-limited.
The plate scale is 0.055\arcsec~pix$^{-1}$.
\label{fig-kspec}}

\figcaption{The sensitivity, $\Delta$K, versus separation as determined
using HD 28867W as a point spread function calibrator.  We should have easily
detected any companion to HD~28867E that had a separation and brightness lying
below the sensitivity curve.
\label{fig-ksens}}

\figcaption{Near-IR spectra obtained with SPEX. The upper trace is the East
component; the middle trace is the West component. The lowest trace is
the South component, an M star. 
The flux scale is absolute, except that the flux of the South component has
been increased by a factor of 10 for display purposes. The East component
is significantly brighter than the West component in the near-IR. 
We have not corrected for telluric aborption.
\label{fig-sp}}  

\figcaption{Near-IR flux ratios. The upper trace is the ratio of the flux in
the East component to that in the West component. There is a clear cool excess
relative to the B9 continuum. The apparent emission in the H~I Brackett
and Paschen lines is attributable to continuum veiling. The lower trace is
the ratio of the East component to the composite B9+G2 spectrum described in
the text. The continuua are well matched, as are most spectral features.
\label{fig-sprat}}  

\figcaption{The E/W flux ratio in the K band. Prominent absorption features
are indicated. The H~I Br~$\gamma$ emision is an artifact of the division,
but the absorption lines are real features. The CO bandheads and the 
neutral absorption are consistent with a G star spectrum.
\label{fig-expsprat}}

\figcaption{Hertzsprung-Russell diagrams showing the allowable locations of
the cool companion. 
The vertical bars mark, from left to right, 
the inferred position for the East B~star 
1500, 1000, 750, 500, 250, and 0~K cooler than the West component.
The vertical extent of the allowable region reflects the $\pm$23~pc
uncertainty in distance from the Hipparcos parallax. The open circle 
denotes the location of HD~283572.
The tracks in the lower plot are from Baraffe et al.\/ (2002); those in the
upper plot are from D'Antona \& Mazzitelli (1994), using mixing length
convection. Baraffe et al.\/ do not provide models for more massive stars.
The isochrones are not directly comparable at ages less than a few million
years. Evolutionary tracks are labelled 
in solar masses; isochrones are labeled in units of 10$^6$~years.
Although the inferred masses and ages differ, it is clear that the
cool companion is young and fairly massive.
\label{fig-hrd}}  

\clearpage
\begin{deluxetable}{lllrrl}
\tablecolumns{6}
\tablewidth{0pt}
\tablecaption{J2000 Positions of HD 28867\label{tbl-optpos}}
\tablehead{
\colhead{Star} & \colhead{RA} & \colhead{DEC} & \colhead{V} & source}
\startdata                            
HD 28867E   & 4 33 33.049  & 18 01 00.20  & 6.99 & Hipparcos\\ 
HD 28867W   & 4 33 32.835  & 18 01 00.58  & 7.04 & Hipparcos\\
HD 28867S   & 4 33 32.806  & 18 00 43.61  & \\
radio source & 4 33 33.023 (.002)  & 18 01 00.47 (.02) & & White et al. 1992\\
X-ray source (E)& 4 33 33.06 (.01) & 18 01 00.3  (.1)  & & Chandra HRC-I\\ 
X-ray source (S)& 4 33 32.81 (.01) & 18 00 43.5  (.1)  & & Chandra HRC-I\\ 
\enddata
\end{deluxetable}

\clearpage
\begin{deluxetable}{lrrrr}
\tablecolumns{5}
\tablewidth{0pt}
\tablecaption{near-IR Photometry\label{tbl-irphot}}
\tablehead{
\colhead{Star} & \colhead{Aperture} & \colhead{$K$} & \colhead{$J-K$} & \colhead{$H-K$}\\
               & \colhead{(\arcsec)}} 
\startdata                            
HD 28867W   &  1.9 & 6.81 $\pm$0.02 & 0.04 $\pm$ 0.03 & 0.08 $\pm$ 0.04\\ 
HD 28867E   &  1.9 & 6.17 $\pm$0.02 & 0.22 $\pm$ 0.03 & 0.12 $\pm$ 0.04\\ 
HD 28867 E+W & 12.6 & 5.69 $\pm$0.02 & 0.12 $\pm$ 0.03 & 0.09 $\pm$ 0.04\\
HD 28867S   &  7.8 & 9.64 $\pm$0.05 & 0.91 $\pm$ 0.08 & 0.20 $\pm$ 0.07\\
\enddata
\end{deluxetable}

\clearpage
\begin{deluxetable}{llrrrl}
\tablecolumns{6}
\tablewidth{0pt}
\tablecaption{IUE obervations\label{tbl-iue}}
\tablehead{
\colhead{Star} & \colhead{Camera} & \colhead{Image} & \colhead{Aperture} & \colhead{time} & \colhead{Date}\\
               &                  &                 &   & \colhead{(sec)}} 
\startdata                            
HD 28867 E+W & SWP-HI & 31967 & Large &  1080 & 1987 March 10 \\
HD 28867 E+W & SWP-LO & 31968 & Large &    24 & 1987 March 10\\
HD 28867 E+W & LWP-HI & 11796 & Large &   600 & 1987 March 10\\
HD 28867W   & SWP-HI & 49847 & Small &  9600 & 1994 January 19\\
HD 28867E   & SWP-HI & 49851 & Small & 10200 & 1994 January 20\\
\enddata
\end{deluxetable}

\clearpage
\begin{deluxetable}{lrll}
\tablecolumns{5}
\tablewidth{0pt}
\tablecaption{Narrow Optical Absorption Features\label{tbl-optabs}}
\tablehead{
\colhead{Observed} & W$_\lambda$ & \colhead{ID} & \colhead{True} \\
\colhead{Wavelength} & (m\AA)     & & \colhead{Wavelength} } 
\startdata            
5447.2 & 118 & Fe I & 5446.575/5446.916 \\
5456.3 &  56 & Fe I & 5455.441/5455.609 \\
5463.7 &  55 & Fe I & 5463.271 \\
5477.5 &  60 & Fe I/Ni I & 5476.287/5476.563/5476.900\\
5481.9 &  31 & Fe I/Ti I & 5480.864/5481.242/5481.429/5481.438\\
5488.7 &  33 & Fe I & 5487.763 \\
5498.1 &  36 & Fe I & 5497.516 \\
5507.2 &  44 & Fe I & 5506.778 \\
5513.2 &  32 & Ca I & 5512.980 \\
5529.0 &  33 & Mg I & 5528.405 \\ 
5545.8 &  76 & Fe I & 5543.147/5543.937 + 5546.500/5546.991\\
5555.1 &  34 & Fe I & 5553.594/5554.882\\ 
5570.2 &  35 & Fe I & 5569.618\\
5573.5 &  51 & Fe I & 5572.841 \\
5576.8 &  31 & Fe I & 5576.090 \\
5582.7 &  25 & Ca I & 5581.965 \\
5587.5 &  45 & Fe I & 5586.756 \\
5595.0 &  37 & Ca I/Fe I & 5594.462/5594.654 \\
5599.1 &  30 & Fe I/Ca I & 5598.289/5598.480 \\
5603.5 &  48 & Fe I/Ca I &  5602.767/5602.842/5602.945\\
5616.1 &  58 & Fe I & 5615.644\\
5625.2 &  42 & Fe I & 5624.038/5624.542\\
5641.9 &  31 & Fe I & 5641.436 \\
5655.6 &  46 & Fe I & 5655.183/5655.489 \\
5659.1 &  65 & Fe I & 5658.531/5658.816 \\
5683.3 &  52 & Ni I/Na I & 5682.198/5682.633 \\
5712.5 &  34 & Fe I/Ni I & 5711.850/5711.883/5712.131 \\
5753.9 &  62 & Fe I/Ni I & 5752.023/5753.121/5754.655 \\
5763.4 &  34 & Fe I & 5762.990 \\
5816.8 &  23 & Fe I & 5816.367 \\
5858.2 &  34 & Ca I & 5857.451 \\
6008.8 &  39 & Fe I & 6008.554 \\
6066.0 &  30 & Fe I & 6065.482 \\ 
6079.2 &  30 & Fe I & 6078.491/6078.999 \\
6103.4 &  55 & Fe I/Ca I & 6102.173/6102.723/6103.186\\
6122.6 &  48 & Ca I & 6122.217 \\
6137.6 &  82 & Fe I & 6136.615/6136.993/6137.694\\
6142.4 &  41 & Fe I & 6141.730\\
6162.7 &  73 & Ca I & 6162.173 \\ 
6170.2 &  62 & Ca I/Fe I & 6169.042/6169.563/6170.504\\
6191.7 &  42 & Ni I/Fe I & 6191.171/6191.558 \\   
6220.0 &  47 & Fe I & 6219.279 \\
6231.3 &  57 & Fe I & 6230.726 \\
6247.0 &  61 & Fe I & 6246.317 \\
6259.1 &  23 & Ti I & 6258.709 \\
6265.0 &  52 & Fe I & 6265.131 \\
6336.9 &  70 & Fe I & 6336.823 \\
6394.2 &  32 & Fe I & 6393.602\\
6400.7 &  71 & Fe I & 6400.000\\
6408.4 &  41 & Fe I & 6408.016 \\
6412.1 &  36 & Fe I & 6411.647 \\
6421.3 &  53 & Fe I & 6421.349 \\
6439.6 &  41 & Ca I & 6439.075\\
6450.4 &  54 & Ca I & 6449.808\\
6456.7 &  41 & Ca I & 6455.598\\
6463.2 &  48 & Ca I & 6462.567\\
6496.0 & 290 & Ca I/Fe I & $\lambda$6495\AA\ blend \\
6500.0 &  43 & Ca I & 6499.650 \\
6547.1 &  43 & Fe I & 6546.245 \\   
6594.0 &  21 & Fe I & 6593.871 \\
6634.3 &  41 & Fe I & 6633.746\\
6644.2 &  50 & Ni I & 6643.629 \\
6664.0 &  91 & Fe I & 6663.5 \\
6678.6 &  47 & Fe I & 6677.909 \\ 
6708.4 &  75 & Li I & 6707.8  \\
6718.0 &  33 & Ca I & 6717.681\\
6734.1 &  52 & Fe I & 6733.151\\
6821.0 &  68 & Fe I & 6820.369 \\
6843.8 & 104 & Fe I & 6842.679, 6843.648\\
\enddata
\end{deluxetable}

\begin{deluxetable}{lllrrrrl}
\tablecolumns{8}
\tablewidth{0pt}
\tablecaption{Companion Parameters\label{tbl-comp}}
\tablehead{
\colhead{T$_{B star}$} & \colhead{SpT} & \colhead{radius} & V & K & \colhead{$\frac{\rm F_B}{\rm F_{comp}}$} & \colhead{$\frac{\rm F_B}{\rm F_{comp}}$} & \colhead{IR standard}\\
\colhead{(K)} & \colhead{} & \colhead{(R$_\odot$)} & \colhead{(mag)} & \colhead{(mag)} & \colhead{(at R$_C$)}& \colhead{(at K)} & \colhead{} }
\startdata                            
10,000    & G1    & 2.6 & 8.4 & 6.6 & 2.6 & 0.69 & HD 27836 (G1V) \\
10,500    & G2-G5 & 2.7 & 8.8 & 6.7 & 3.7 & 0.83 & HD 6582 (G5V) \\
10,750    & G5-K0 & 2.8 & 8.7 & 6.8 & 3.6 & 0.92 & ---  \\
11,000    & K0-K2 & 2.9 & 9.1 & 6.8 & 4.9 & 1.00 & HD 112758 (K0V) \\
11,250    & K2-K5 & 3.2 & 9.6 & 6.9 & 7.6 & 1.11 & HD 4628 (K2V) \\
11,500    & K5-K7 & 3.4 &10.1 & 6.8 &10.9 & 1.06 & HD 97503 (K5V) \\
\enddata
\end{deluxetable}

 
\begin{figure}
\plotone{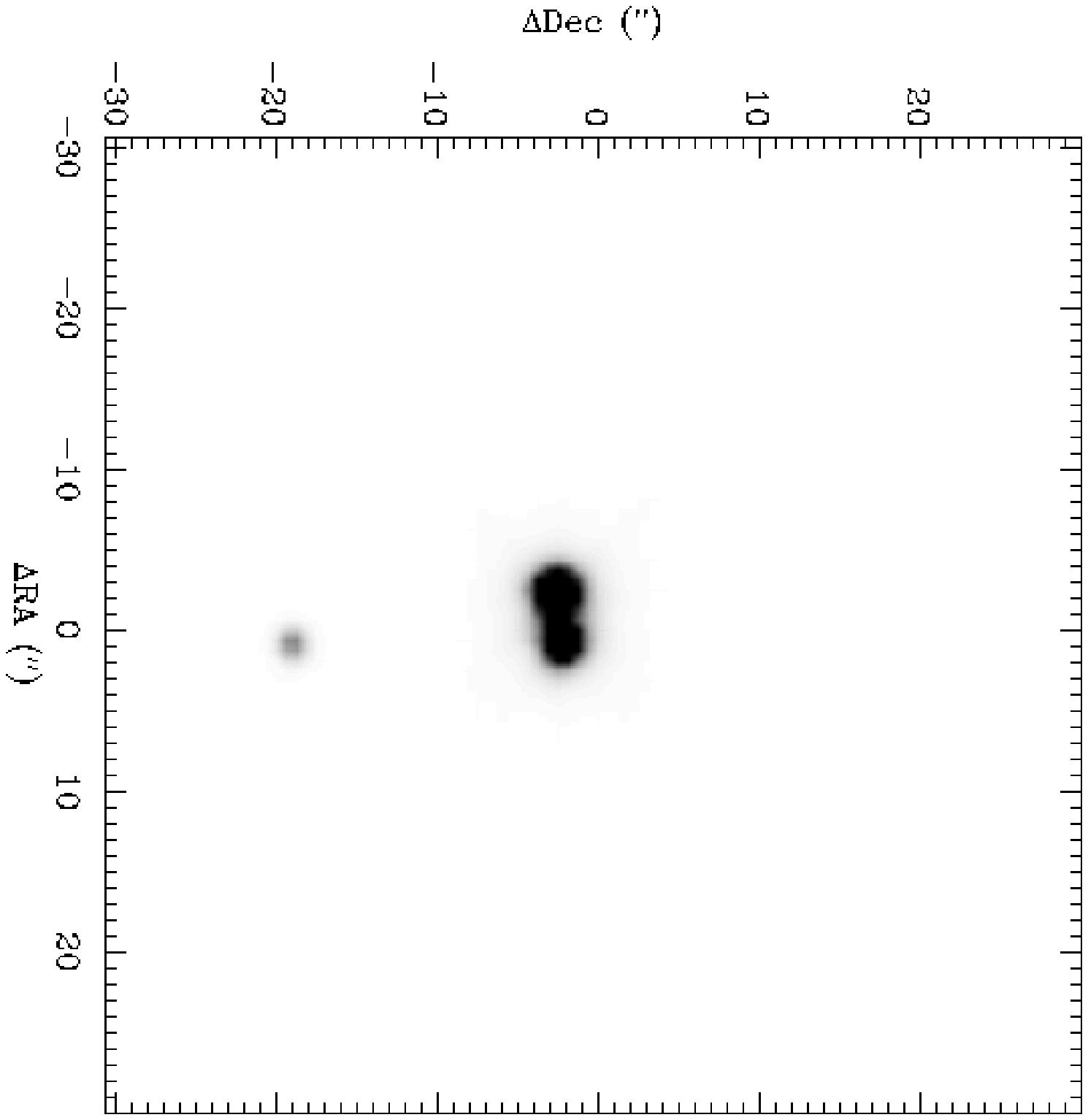}
\end{figure}
 
\begin{figure}
\plotone{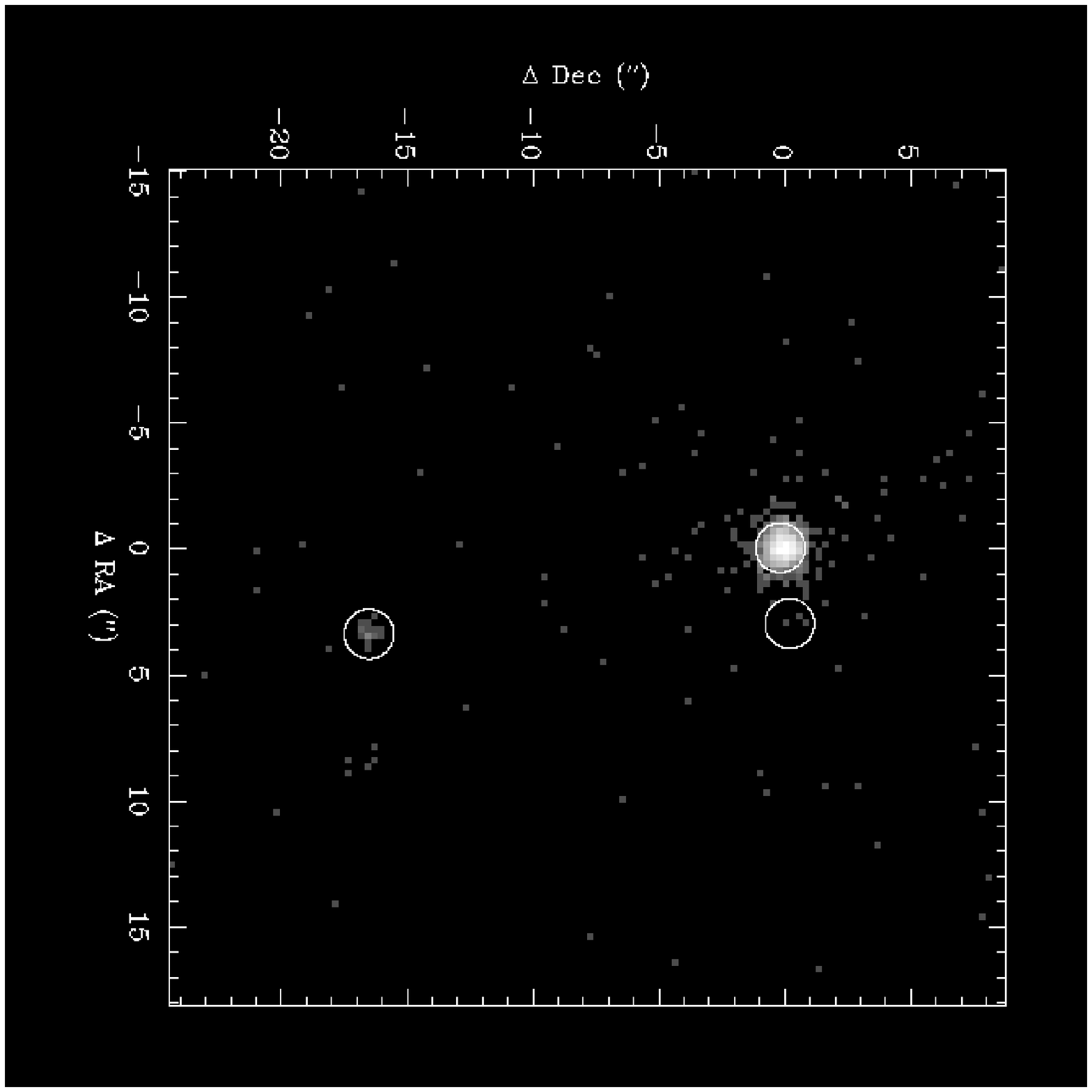}
\end{figure}

\begin{figure}
\plotone{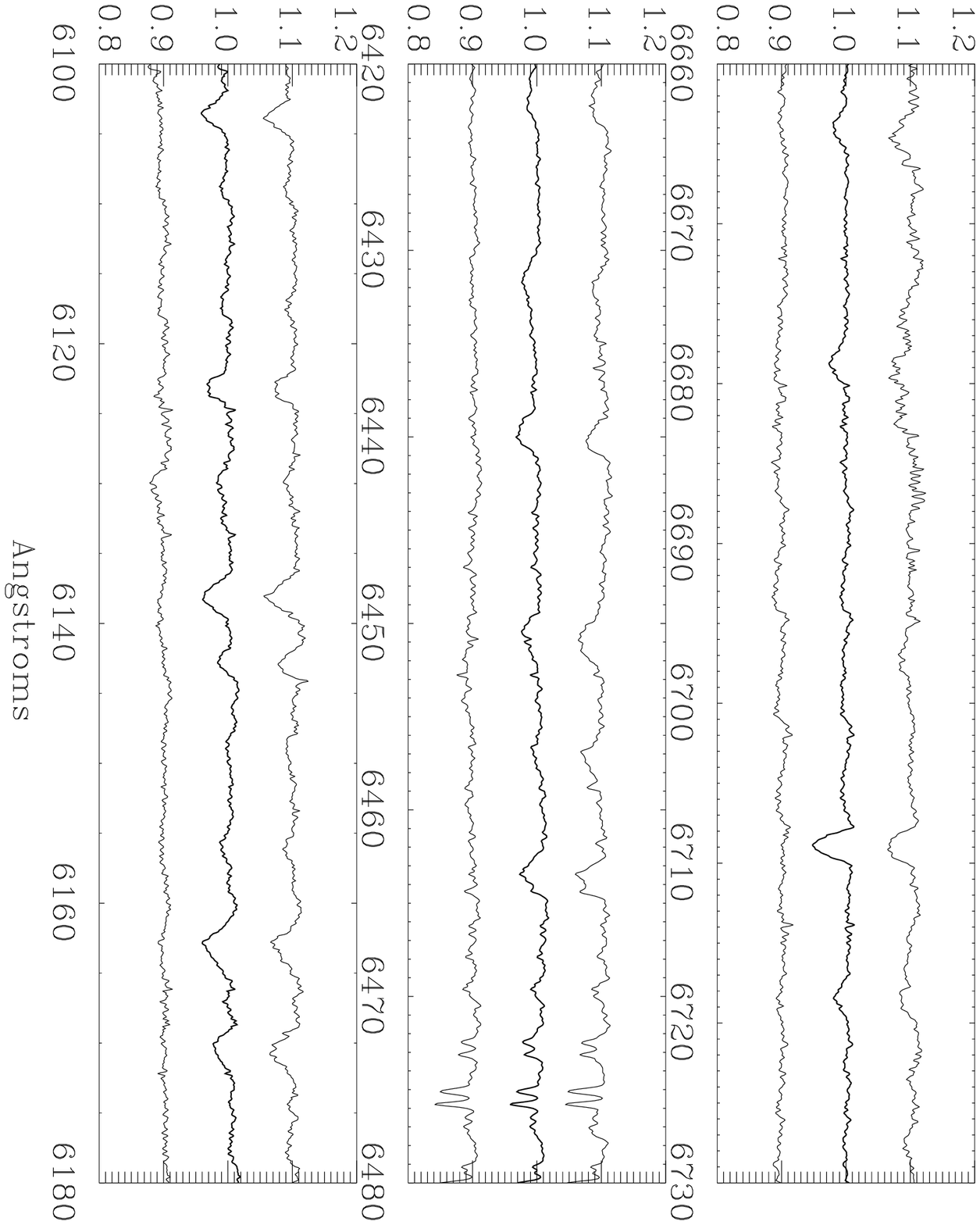}
\end{figure}

\begin{figure}
\plotone{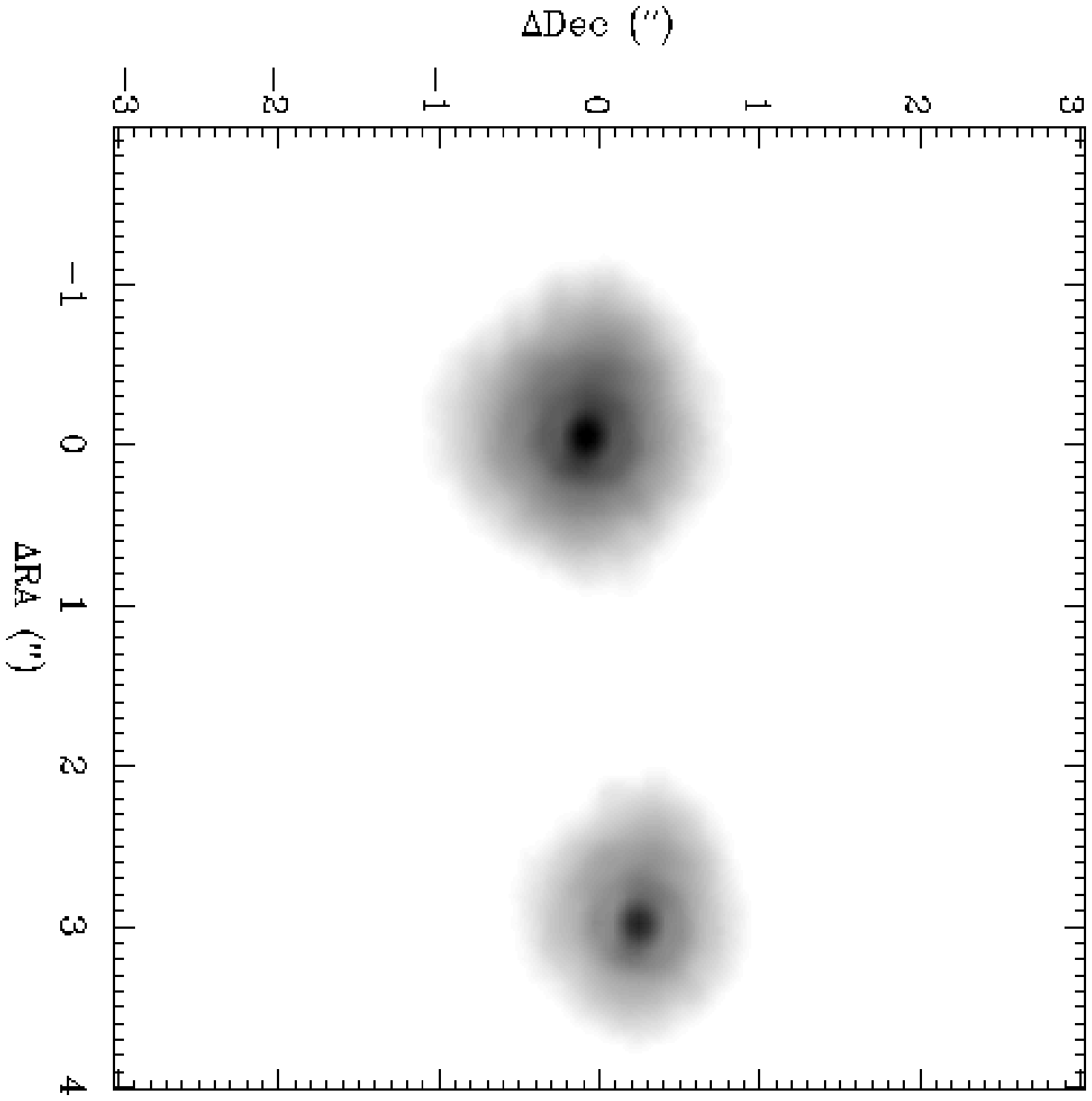}
\end{figure}

\begin{figure}
\plotone{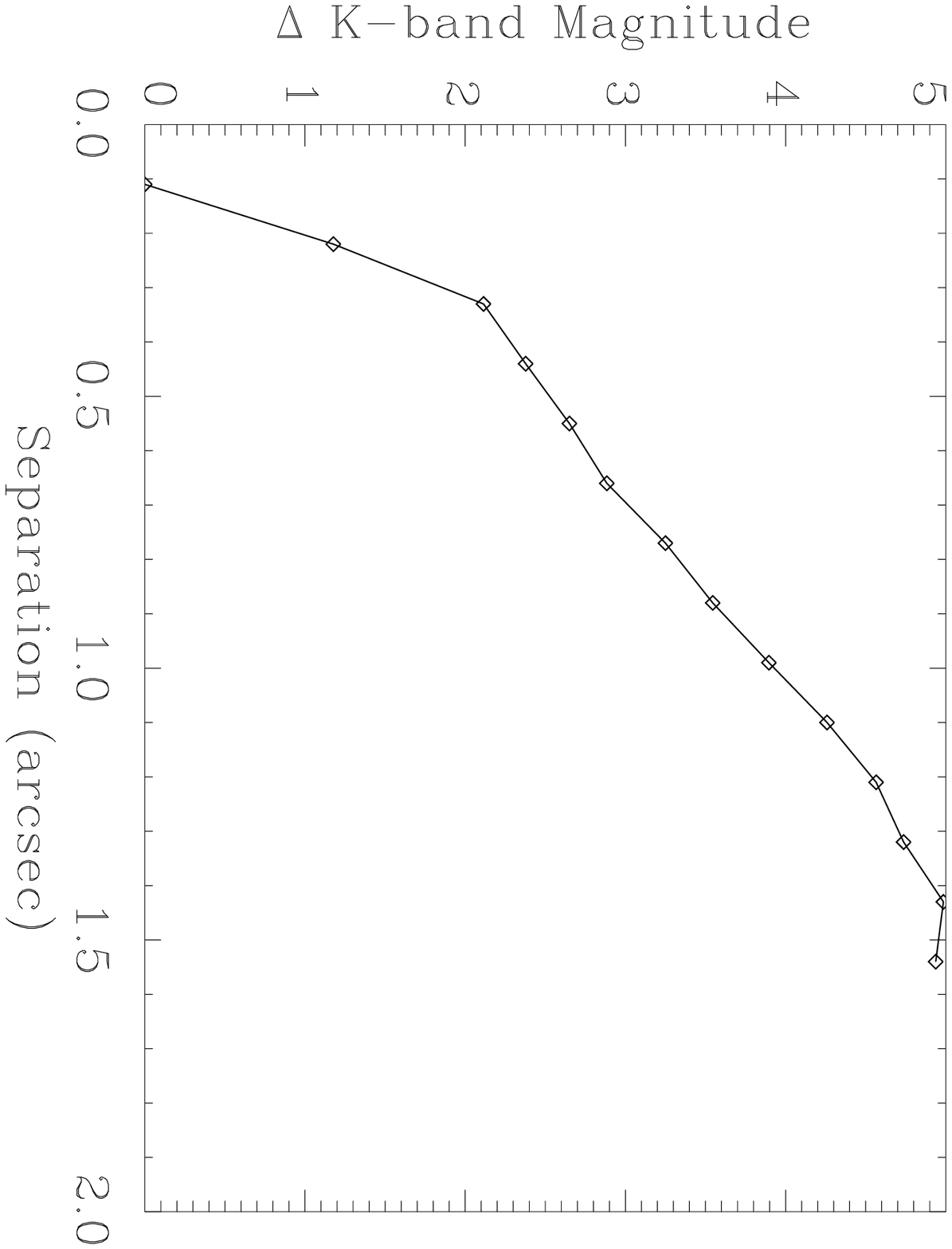}
\end{figure}

\begin{figure}
\plotone{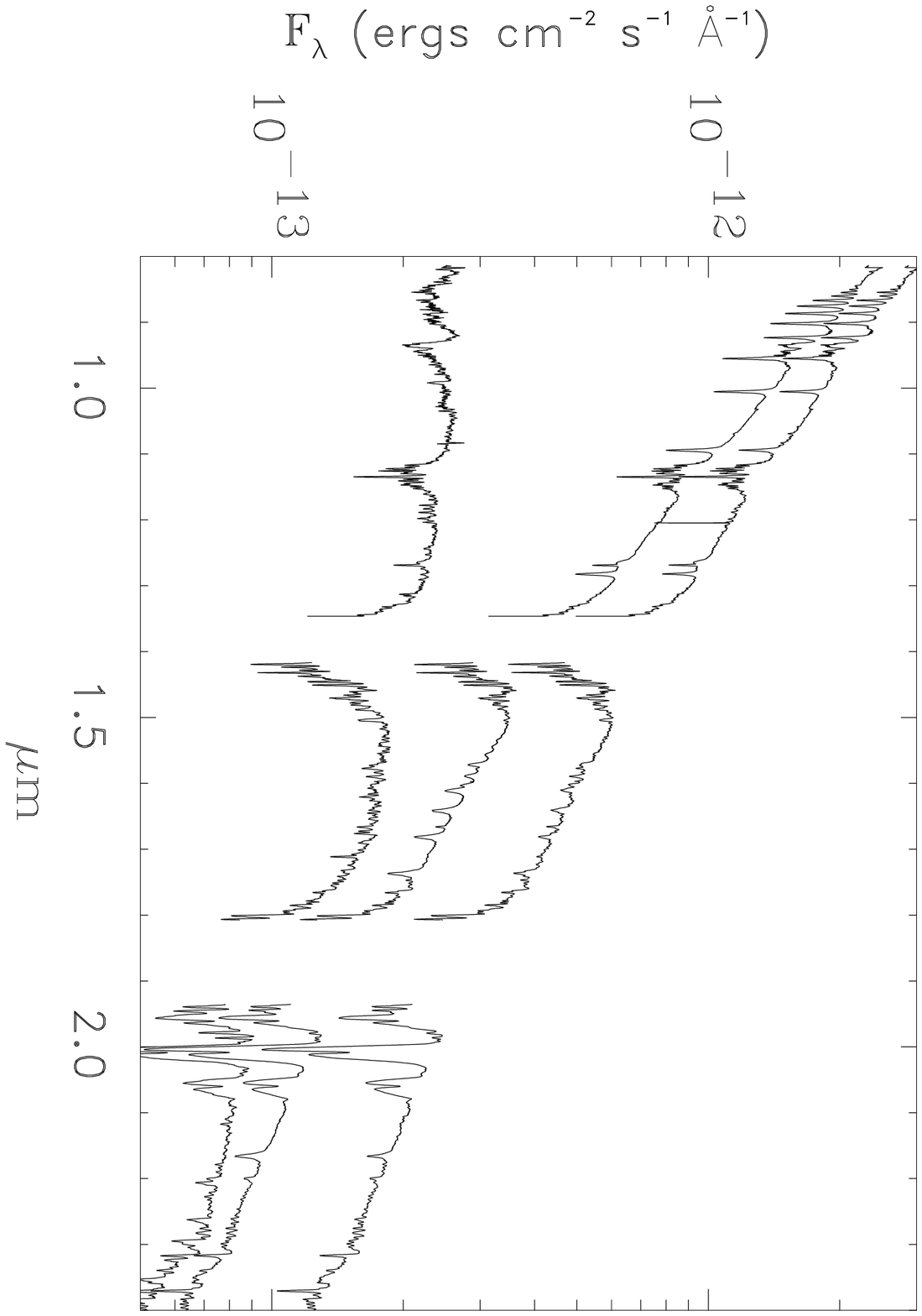}   
\end{figure}

\begin{figure}
\plotone{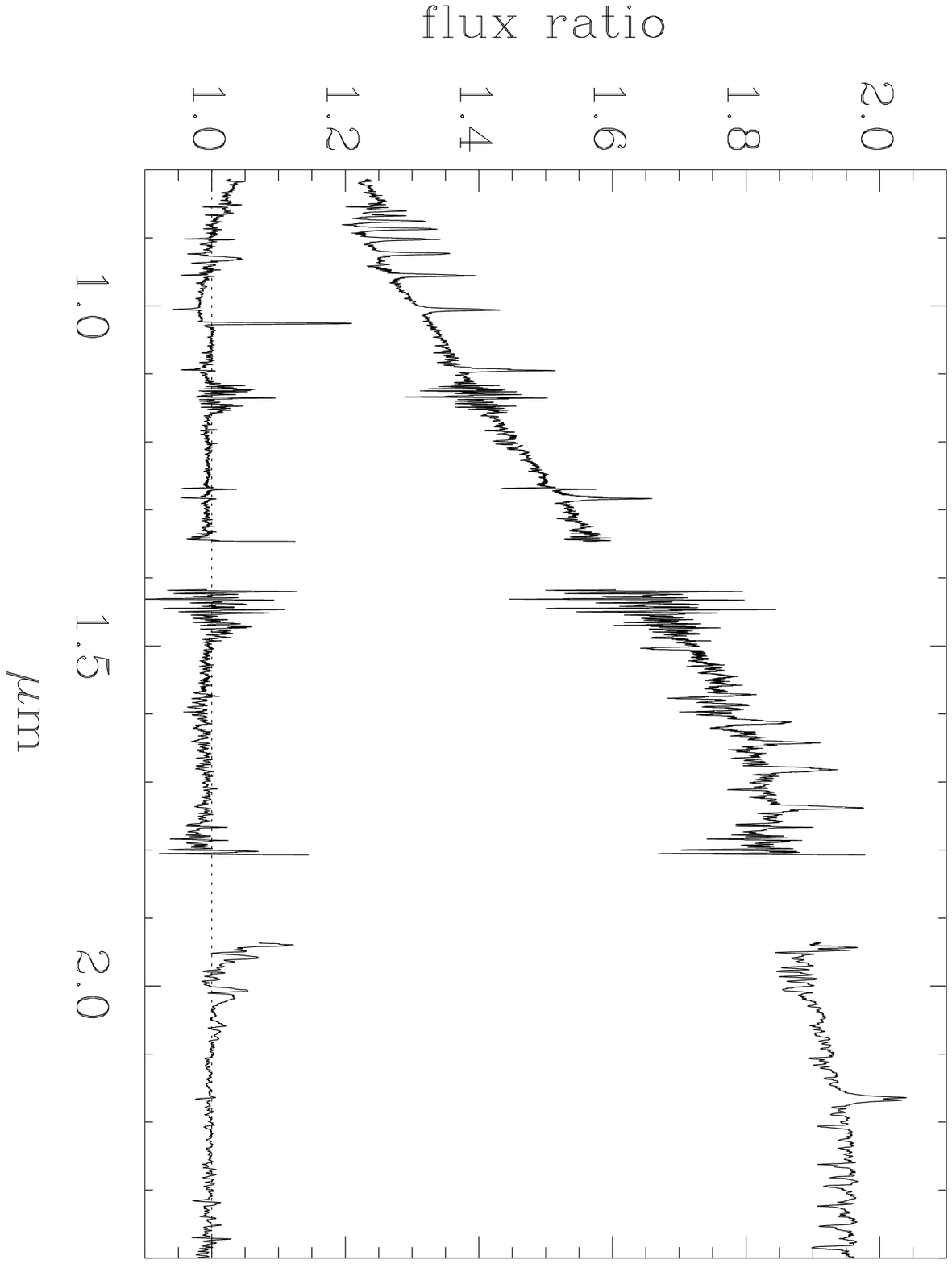}
\end{figure}

\begin{figure}
\plotone{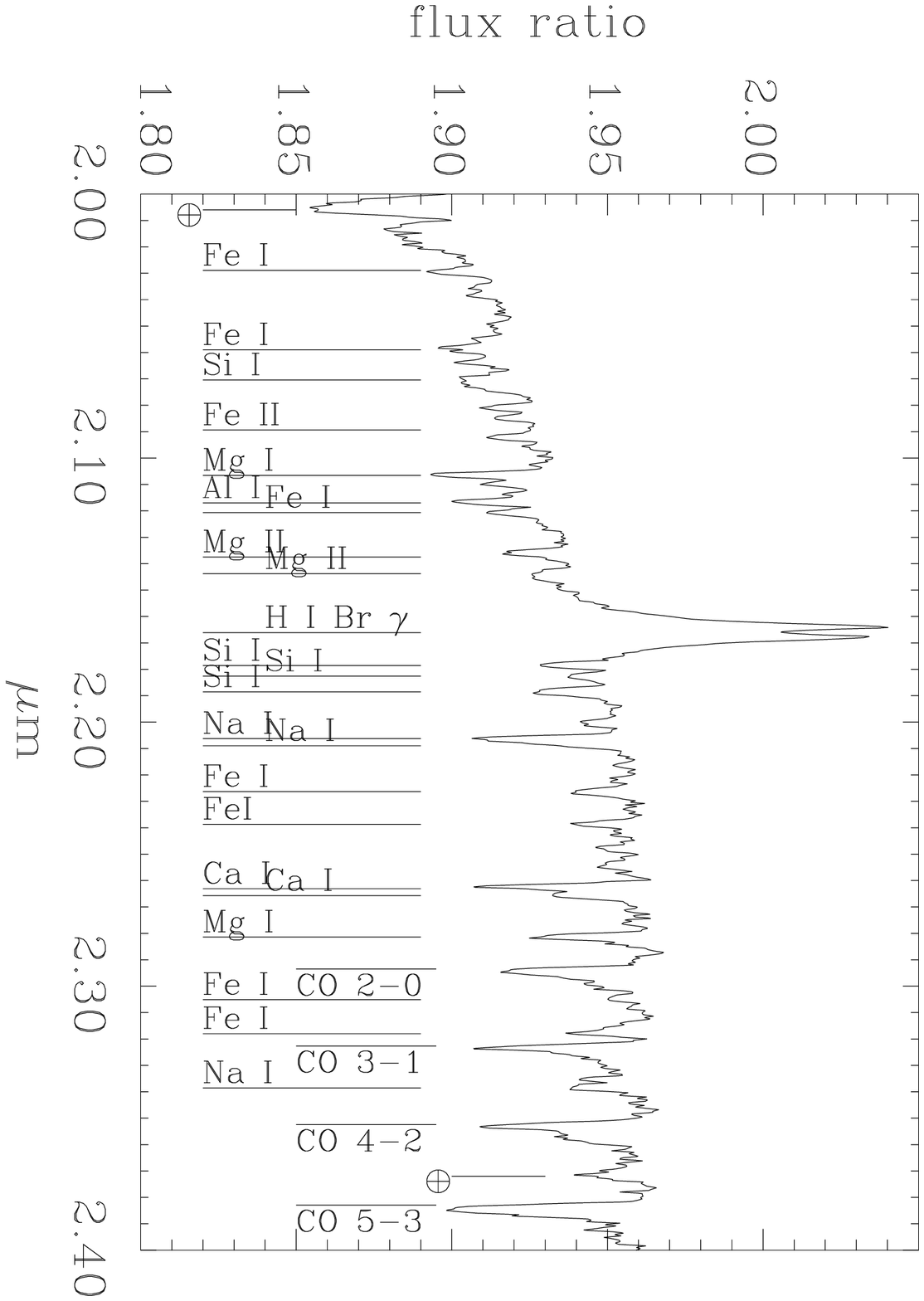}
\end{figure}

\begin{figure}
\plotone{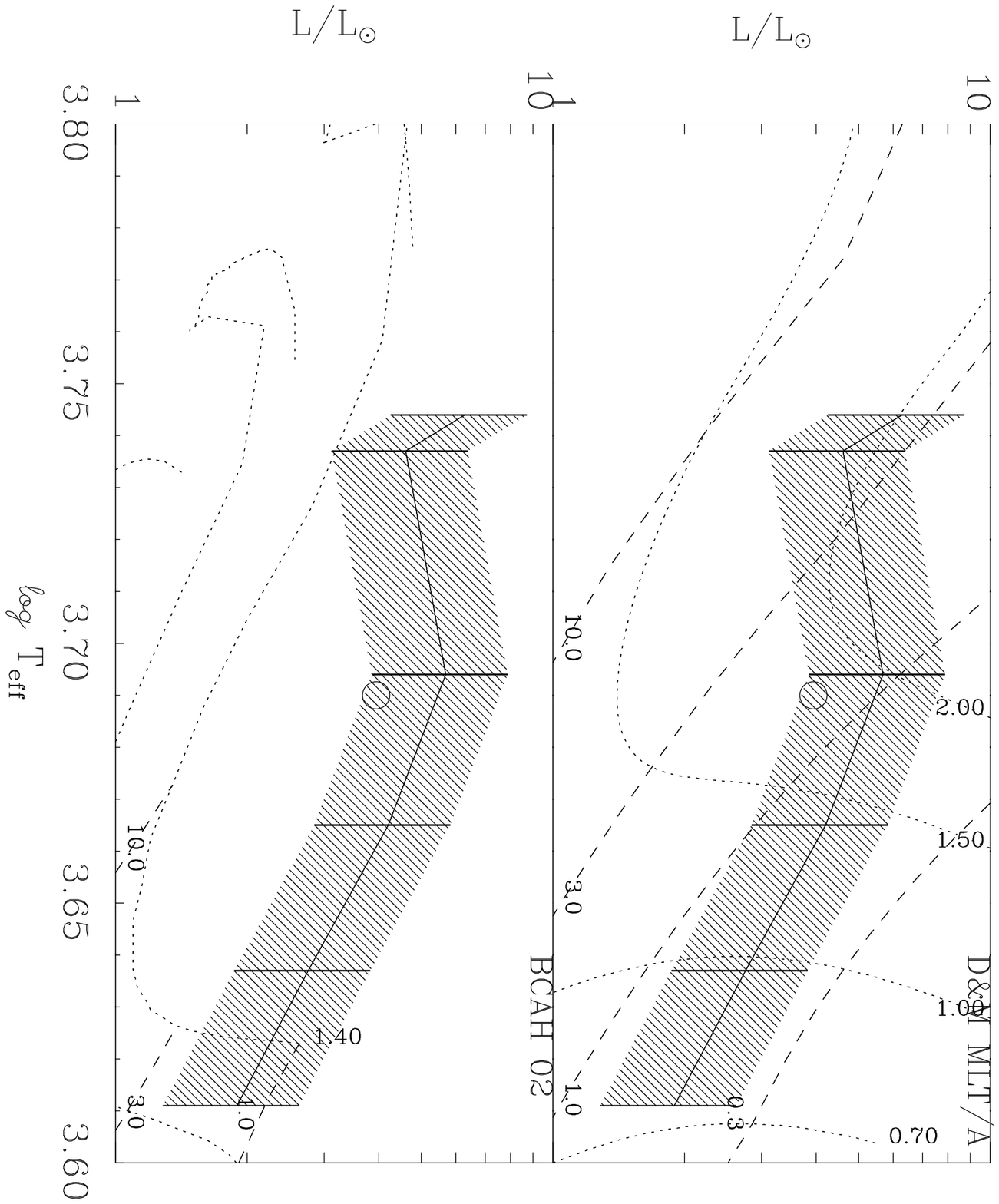}
\end{figure}

\end{document}